\documentclass[useAMS,prd,aps,floats,twocolumn]{revtex4}

\usepackage{graphicx}
\usepackage{axodraw}
\usepackage{color}
\usepackage{psfrag}

\def\refe@jnl#1{{#1}}
\def\aj{\refe@jnl{Astron.~J.}}
\def\araa{\refe@jnl{Annu.~Rev.~Astron.~Astrophys.}}
\def\apj{\refe@jnl{Astrophys.~J.}}
\def\apjl{\refe@jnl{Astrophys.~J.~Lett.}}
\def\aap{\refe@jnl{Astron.~Astrophys.}}
\def\mnras{\refe@jnl{Mon.~Not.~R.~Astron.~Soc.}}
\def\prd{\refe@jnl{Phys.~Rev.~D}}
\def\fcp{\refe@jnl{Fund.~Cos.~Phys.}}
\def\physrep{\refe@jnl{Phys.~Rep.}}
\def\physlett{\refe@jnl{Phys.~Lett.}}

\def\mdm{{m_{\rm dm}}}

\begin{document}

\title{ Light Dark Matter Annihilations into Two Photons}

\author{C. Boehm\footnote{
On leave LAPTH, UMR 5108, 9 chemin de Bellevue - BP 110, 74941 Annecy-Le-Vieux, France.} \footnote[3]{ preprint
numbers: LAPTH--1156/06; PCCF RI 0603.} }

 \affiliation{ Physics department, Theory Unit, CERN, CH-1211 Gen{\`e}ve 23,
  Switzerland}

\author{J. Orloff}
\affiliation{LPC, Universit{\'e} Blaise Pascal, 63177 Aubi{\`e}re Cedex, France}
\author{P. Salati}
\affiliation{LAPTH, UMR 5108, 9 chemin de Bellevue - BP 110, 74941
Annecy-Le-Vieux, France}

\begin{abstract}
  We compute the pair annihilation cross section of light (spin-0) dark matter particles
  into two photons and discuss the detectability of the monochromatic line associated with these annihilations.
\end{abstract}

\maketitle


\section{Introduction}
The precise determination by INTEGRAL/SPI \cite{Jean:2003ci} of the characteristics of the 511 keV line emitted in
our galaxy
\cite{Dixon:1997nn,Sreekumar:1997yg,Churazov05,Milne:2001zs,Dermer:1997cq,Pohl:1998bs,Dermer:2001wc,Milne:2001dt}
has shed new light on the physics of the inner part of the Milky Way. This line has now been identified with a
high level of confidence as originating from electron-positron annihilations. Although this recent detection
probes unambiguously the existence of anti-matter inside our galaxy, its origin remains unknown.

The observation of a relatively high fraction of low energy positrons in the bulge and a low fraction in the disk
 certainly constitutes the most puzzling aspect of this emission.

Most of the astrophysical sources that have been proposed in the literature (e.g. Wolf-Rayet stars, Hypernovae,
cosmic rays, pulsars, black holes) are associated with a low value of the bulge-to-disk (B/D) ratio or cannot
explain why the 511 keV radiation seems to follow the stellar morphology of the galactic bulge.

The remaining plausible sources are old galactic populations, such as Low Mass X-ray Binaries (LMXB) and Type 1a
Supernovae (SN1A) \cite{Knodlseder05}.  However, to explain the observed flux and positron distribution, they both
rely on strong hypothesis. LMXB require that the positrons emitted in the disk escape into the bulge while SN1A
need a positron escape fraction and an explosion rate that are large enough to maintain a steady flux.

At least eight point sources could explain the diffuse emission as observed by SPI \cite{Knodlseder05}. However
Ref.~\cite{Weidenspointner:2006nu} did not find any evidence for significant emission from point sources in the
galactic centre as yet.

Another candidate could be light dark matter (LDM) particles \cite{bens,bf,511} annihilating into
electrons--positrons, neutrinos and photons. The positrons thus emitted lose their energy by ionization and
eventually form para-positronium atoms with the thermal electrons present in the bulge of the galaxy \cite{511}.
The 511 keV line emission is expected to be strongly correlated with the dark matter energy density distribution.
The latter is maximal in the inner part of the galaxy so the positrons should mostly be produced in the centre of
the Milky Way and should naturally stop on the electrons present in the bulge. Depending on the cuspyness of the
profile, this would explain why the emission is well described by a sphere of only $\sim$ $8-10^\circ$ of
diameter.

If dark matter is light enough, the amount of low energy gamma rays
produced by the dark matter (DM) annihilations remains compatible
with observations. Below the muon mass threshold ($m_{dm} \leq 100$
MeV), the gamma ray production channels are: e.g. the DM pair
annihilation into electron-positron plus a photon
\cite{bens,beacom,bu}, positronium formation, inflight $e^+ e^-$
annihilations, initial and final state radiation associated with the
electron-positron annihilations (although they have not been
included as yet in previous studies).

All these processes generate a continuum. In addition, lines are produced at an energy corresponding to either the
dark matter mass or at 511 keV. The former is the subject of the present paper. If the corresponding flux is large
enough, this could be an unique
 tracer to answer the question of the low energy positrons.

After summarizing recent progress on this topic and their
implication for the dark matter characteristics, we will estimate
the annihilation cross section of light dark matter particles into
two photons and discuss the observability of the line at
$E_\gamma=m_{dm}$. We base our analysis on the model proposed in
Ref.~\cite{bens,bf}, which has been studied in detail in
Ref.~\cite{boehm,boehmS,ascasibar,boehmascasibar}.

\section{Dark matter characteristics}

\subsection{Dark matter mass}

The initial mass range proposed to fit SPI data was a few MeV to 100 MeV \cite{511}. However, recent analysis now
provide more stringent constraints.

From the estimate of the ($\mbox{dm} \ \mbox{dm}^{\star} \rightarrow e^+ \, e^- \, \gamma$) process,
Ref.~\cite{beacom} found that the dark matter mass should be smaller than 20 MeV.  However the formula that they
used is incorrect within the framework that we consider in this paper. The correct expression, embedded into a
more complete astrophysical analysis (in which INTEGRAL/SPI response function to the 511 keV source and a
modelisation of the background were implemented), gives $m_{dm} \leq$30 MeV \cite{bu}.

From the fraction of ortho and para--positronium, Ref.~\cite{jean} found that the energy injection of the low
energy positrons in our galaxy should be smaller than 10 MeV. Conservatively, this imposes $m_{dm} < 10$ MeV.

From the inflight $e^+ e^-$ annihilations, Ref.~\cite{beacom2} found an upper limit of 3 MeV. This result is based
on two assumptions. Due to similar values of the flux below 1 MeV in the galactic disk at different longitudes,
$(8.32 \pm 1.8) \ 10^{-3}  {\rm ph \; cm^{-2} \; s^{-1} \; sr^{-1} \; MeV^{-1}}$ for $330<l<30^\circ$ and $(7.87
 \pm 2.5) \ 10^{-3}\ {\rm ph \; cm^{-2} \; s^{-1} \; sr^{-1} \; MeV^{-1}}$ for $330<l<350^\circ$,
 $10<l<30^\circ$ ($|b|<10^\circ$, that is the galactic disk region),
the authors assumed that the diffuse background had the same \textit{astrophysical} origin in the inner and outer
part of the galaxy. Also they assumed that one could extrapolate the power law description of the continuum at low
energy to higher energy. With these assumptions, their estimate of the inflight annihilations at high energy is
1.2 to 1.3 larger than their modelisation of the continuum background. Such a bound is certainly very stringent.
However, for the moment, we prefer to be cautious since they did not include all the processes that are
responsible for the positron cooling and, perhaps even more importantly, the validity of their two main hypotheses
needs to be demonstrated.

Finally, let us mention that  an upper bound of 7 MeV has also been found in Ref.~\cite{boehmascasibar,ascasibar}
from the estimate of the electron $g-2$.

Note also that in Ref.~\cite{Serpico:2004nm,SN}, $\mdm$ is constrained to be larger than 2 and 10 MeV from
nucleosynthesis and core collapse supernovae respectively. Ref.~\cite{Serpico:2004nm} conclusions were obtained by
assuming that DM was in thermal equilibrium at nucleosynthesis. However, for the cross section we consider, the
neutrino thermal decoupling temperature is larger than 1 MeV. Also Ref.~\cite{SN} concludes that $\mdm > 10$ MeV
if the main annihilation channel is into neutrino pairs in the primordial universe. However, in our case, dark
matter annihilates predominantly into electron-positron \cite{boehm}.

All these bounds are based on assumptions. To understand how the line flux scales with the dark matter mass, we
will consider values of the dark matter mass ranging from the electron mass $m_e$ to  100 MeV.  The reader is
nevertheless invited to keep in mind that large values of the dark matter mass are disfavoured.

\subsection{Dark matter density profile}

Due to its coded mask, SPI observation cannot be related to a unique source. In particular, it is not possible to
discriminate between low energy positrons originating from either dark matter annihilations or a set of
astrophysical sources.

The hypothesis of a dark matter source (and the response to it by SPI) has been studied in detail in
Ref.~\cite{ascasibar}. The only dark matter scenario which can explain SPI observation requires a
Navarro-Frenk-White profile (NFW) and a velocity-independent cross section. More precisely, using the conventional
parametrization of the profile, the parameter $\gamma$ (which fixes the slope of the profile in the inner part of
the galaxy) was found to be equal to $1.03 \pm 0.04$, assuming $r_0=16.7$ kpc and $\rho_0=0.347$ GeV/cm$^3$
\footnote{These parameters correspond to the requirement of a local dark matter energy density of 0.3 GeV/cm$^3$
in the solar neighborhood, a virial radius $R_{vir} \approx 260$ kpc and a mass $M_{vir} \approx \, 10^{12}
M_{\odot}$.}.

The authors tested different profiles, ranging from flat to cuspy (the cuspiest corresponds to that in
Ref.~\cite{Moore99}, hereafter denoted M99). They tested decaying dark matter particles (as proposed in e.g.
Ref.~\cite{hooper,pospelov,yanagida}) as well as annihilating particles with a pure velocity-dependent and/or
independent annihilation cross section \cite{bf}.

The Maximum Likelihood Ratio (MLR) was obtained for a NFW model
together with a velocity-independent cross section, excluding all
other types of light dark matter scenarios. This conclusion might be
alleviated if one considers extremely cuspy profiles. Indeed the MLR
for scenarios based on velocity-dependent cross section or decaying
particles increases with the cuspyness of the profile. Nevertheless,
even for M99, the fit associated with a velocity-dependent cross
section remains very bad. The fit is even worse for decaying
particles. Hence, for conventional profiles as displayed in the
literature, velocity-dependent cross section and decaying particles
scenarios are excluded as a possible explanation for the 511 keV
line.

To revive a pure velocity-dependent cross section,  the profile would have to satisfy:
\begin{equation}
\label{v/r}  \frac{v}{(r/r_0)^{\xi}} \sim \sqrt{\frac{\sigma_a}{\sigma_b}}
\end{equation}
where $\xi = \gamma-1$ and with $v$ growing as $r^{\delta}$ ($\delta>1$ for M99 while, for NFW, $\delta>1.5$ in
the inner kpc). Here $\sigma_b \sim 10^{-25}$ cm$^3$/s corresponds to the value of the annihilation cross section
that satisfies the relic density condition (the precise value depends on the scenario) while $\sigma_a \sim 2.6 \
10^{-30}$ cm$^3$/s is the value required to fit INTEGRAL/SPI data (see next subsection). The condition expressed
in Eq.~\ref{v/r} thus imposes the relationship $\gamma = \delta +1$ between the density and velocity dispersion
profiles and also requires $r_0 \sim \left(\frac{\sigma_a}{\sigma_b}\right)^{1/(2 \delta)}$. This suggests that,
for ``reasonable'' values of $r_0$, the density and velocity dispersion profiles must for example behave as
$r^{-3}$ and $r^2$ respectively. Whether these behaviours can be reproduced by using numerical simulations and can
agree with observations is another matter.

Eventually, the very precise determination of the Milky Way dark matter halo profile could confirm or infirm the
LDM hypothesis. At present, the profile is not known in the inner kpc where the signal comes from.

Simulations of galactic dark halos favour cuspy profiles (i.e. with a radial dependence given by $\rho \propto
r^{-\gamma}$ where $\gamma \sim 1$) but they still lack of resolution at the scale of interest. Besides,
generally, these simulations only give an information of the shape of the profile  (often assuming sphericity) at
the epoch of formation and do not take into account astrophysical processes that could perhaps affect the shape
and the radial dependence.

 Observations are
therefore expected to be more powerful but they are more controversial. From the value of the optical depth
($\tau$) of our galaxy, as given by MACHO in 2001 \cite{MACHO}, Ref.~\cite{be} concluded that the profile should
behave as $r^{-0.3}$. However recent $\tau$ measurements \cite{Ansari:2004fg,griest} now favour cuspy profiles. In
fact, Ref.~\cite{Battaglia:2005rj} even concluded that a NFW profile with a concentration parameter $c=18$
reproduces -- as a whole-- the trends in the data better than a truncated flat model (with $\rho \propto r^{-2}$
in the inner part) although the latter cannot be excluded (it provides a better fit at larger radii where the
profile behaves as $\rho \propto r^{-5}$).

These results disagree with some of the observed profiles of dwarf galaxies which are shallower than predicted
\cite{gentile}. However observations and predictions can perhaps be reconciled if one takes into account e.g.
triaxial effects, astrophysical processes, the fact that some dwarfs are being disrupted and/or the presence of a
bar \cite{hayashi,Rhee}.

\subsection{Tree-level annihilation cross section to fit the 511 keV line}
According to Ref.~\cite{ascasibar}, the LDM scenario reproduces the
observed flux when
\begin{eqnarray}
\label{sigma511} \sigma_{511} v_r\, &\sim& 2.6 \, 10^{-30} \, \left(\frac{m_{dm}}{\mbox{MeV}}\right)^2 \
\left(\frac{r_0}{16.7 {\rm kpc}}\right)
\nonumber\\
&& \hspace{1cm} \times \ \left(\frac{\rho_0}{0.347 {\rm GeV/cm^3}} \right)^2 \mbox{cm}^3/\mbox{s}.
\end{eqnarray}

With this \textit{velocity-independent} annihilation cross section, LDM particles must be lighter than 30 to 100
MeV to not overproduce low energy gamma rays via final state radiation \cite{ascasibar,bu}.

\subsection{Nature of light dark matter}

As explained above, only scenarios, for which the pair annihilation cross section into $e^+ e^-$ is
velocity-independent, provide a good fit to INTEGRAL/SPI observations. This selects either spin-0 dark matter
particles exchanging a heavy fermion $F_e$, spin-1/2 particles exchanging a heavy scalar $S_e$ or spin-1/2 dark
matter candidates coupled to a new gauge boson with axial couplings to electrons and positrons.

The second possibility is excluded because it requires too small values of the mass of the scalar $S_e$ for
perturbative values of the couplings. The third scenario is also very constrained (if dark matter is axially
coupled to the gauge boson) because the ratio $\sigma_{511} v_{r} / \sigma_{ann} v_{r}$, for $\mdm >1$ MeV, scales
as $7 \ \, m_e^2 \, {m_{dm}}^{-2} \, v^2$ with $v << 10^{-3} c$. This ratio is therefore too small in the inner
kpc to fit SPI data; it should be about $10^{-5}.$

Eq.\ref{sigma511} therefore favours spin-0 dark matter candidates coupled to a heavy fermion $F_e$.

Our model nevertheless contains an additional ingredient. To achieve the correct relic density, one needs to
introduce a new gauge boson. The associated pair annihilation cross section being velocity dependent, it can also
evade low energy gamma ray constraints. On the other hand, it will not contribute significantly to the
monochromatic line emission.

For these reasons, we shall consider only scalar dark matter particles annihilating through the exchange of heavy
fermions. Possibilities to revive the fermionic case will be discussed elsewhere. Note that the order of magnitude
of $ \sigma_{511} v_r$ is comparable to that expected in a supersymmetric scenario. However our model is somewhat
simpler.

\section{Dark matter annihilation cross section into two photons}

In what follows, we consider the Lagrangian
$${\cal L}= \bar\psi_{F_e} (c_r P_L+c_l P_R) \psi_e \phi_{dm} + h.c
$$
where $P_{R,L}$ are the chiral projectors $(1\pm\gamma_5)/2$. The relevant diagrams are box-diagrams containing 1,
2 or 3 heavy fermions $F_e$. Assuming that $dm \neq dm^{\star}$ (which fixes the circulation of arrows), there are
6 diagrams, taking into account permutation of the photon external legs.

\begin{figure}[h]
  \includegraphics[width=7cm]{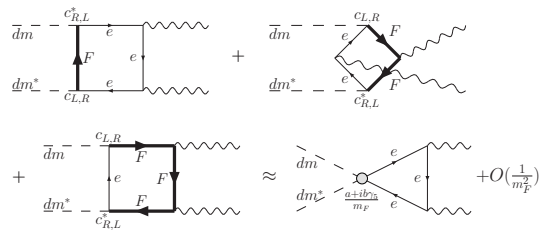}
  \caption{First order contribution to the pair annihilation into two photons for the $F$ exchange.}\label{fig3}
\end{figure}

From naive power counting, each box is logarithmically divergent.
However, gauge invariance dictates a result proportional to
$F_{\mu\nu}^2$ rather than $A_\mu^2$. This requires 2 powers of
external momenta, so that the integrand must in fact converge like
$d^4k/k^6$ for large loop momenta $k$. In the limit $m_{F_e} \gg
m_{e,dm}$ (relevant due to LEP and other collider/accelerator
constraints), the contribution of momenta larger than $m_{F_e}$ is
$\sim 1/m_{F_e}^2$. The leading $1/m_{F_e}$ term can thus be safely
obtained by expanding the integrand in powers of $1/m_{F_e}$ and
keeping only the first term. This corresponds to ``pinching'' the
box with one $F_e$ into a triangle involving only electrons and an
effective dm-dm-e-e coupling given by
$$
{\cal L}_{eff}= \frac{1}{m_{F_e}} \phi_{dm}^*\phi_{dm} \bar\psi_e (a+ib\gamma_5)\psi_e
$$
with the real couplings $a,b$ given by $a+ib=c_l^* c_r$.

For this set-up, computing the cross-section is a loop-textbook exercise
for which we find:
\begin{eqnarray}
\sigma_{\gamma \gamma}v_r &=& \frac{\alpha^2} {(2\pi)^3 \  m_{F_e}^2} \ \frac{m_e^2}{m_{dm}^2} \times \left[ b^2
|2 C_0 m_{dm}^2|^2\right.
\nonumber\\
&&\hskip2em + a^2\left.|1+ 2 C_0 (m_e^2 - m_{dm}^2)|^2\right].
\end{eqnarray}
$C_0$ is a function of $m_e$ and $m_{dm}$ given by the
Passarino-Veltman scalar integral:
\begin{eqnarray}
&&C_0(q_1^2=0,(q_1+q_2)^2=4m_{dm}^2,q_2^2=0,m_e^2,m_e^2,m_e^2) \nonumber\\
\nonumber\\
&&=\int \frac{d^4 k}{i \pi^2}  \frac{1}{\left((k+q_1)^2-m_e^2\right)
(k^2-m_e^2)\left((k-q_2)^2-m_e^2\right)}\nonumber\\
&&=\frac {1} {4m_{dm}^2}\int_0^1 \frac{dx}{x} \ln(1-x(1-x)4m_{dm}^2/m_e^2
-i\varepsilon_F)
\nonumber
\end{eqnarray}
with $q_{1,2}$ the external photons four-momenta.  For $m_{dm}>m_e$,
this function develops an imaginary part corresponding to the
formation of a real $e^+e^-$ pair subsequently annihilating into 2
photons, and giving the largest contribution for masses above 1 MeV.

For $m_{dm}\ll m_e$, $C_0$ behaves as $[-1/(2m_e^2)+m_{dm}^2/(3m_e^4)]$, so that both terms of the cross section
behave as $m_{dm}^2/(m_e m_{F_e})^2$. This limit is relevant to estimate the effect of heavier particles than the
electron in the loop. For example, the contribution of the $\tau$ lepton could be significant if the corresponding
couplings $(a_\tau,b_\tau)$ are larger than $\approx (m_{\tau}/m_{dm}) \times (a_e,b_e) \times
(m_{F_{\tau}}/m_{F_e})$ (with $\mdm < m_{\tau}$), i.e. if they scale at least like usual Yukawa couplings. Since
an independent detailed analysis is required to check whether or not such couplings can pass particle physics
constraints, we prefer giving a conservative estimate based on the electron contribution only. The latter cannot
be turned off without losing the 511 keV line signal. It therefore constitutes a safe lower bound on the
detectability of the line at $E_\gamma=m_{dm}$.

Within the pinch approximation, the cross-section relevant for the
origin of the 511 keV emission is:
$$
\sigma_{511}v_r= \frac{\beta_e} {4 \pi m_{F_e}^2} \left(a^2 \beta_e^2 + b^2 \right)
$$
with $\beta_e = \sqrt{1-m_e^2/m_{dm}^2}$, which indeed for $b=0$ reduces to the expression used e.g. in
Ref.~\cite{ascasibar} for large $m_{F_e}$. After careful comparison with SPI data, the last reference found
$$
\sigma_{511} v_r = 2.6 \ 10^{-30} \ \left(
\frac{m_{dm}}{\mbox{MeV}}\right)^2
 \mbox{cm}^3/\mbox{s}.
$$

\begin{figure}
  \includegraphics[width=7cm]{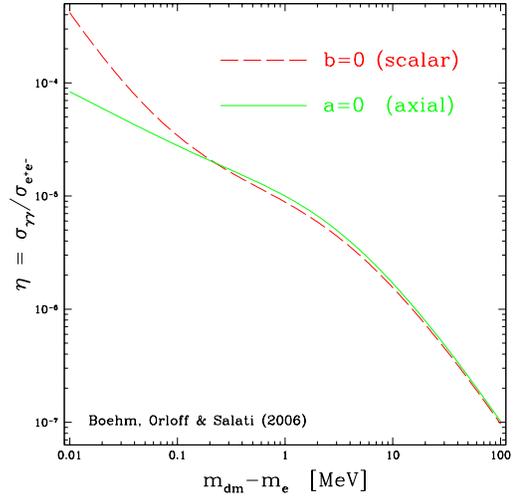}
  \caption{Ratio $\eta$ of annihilation cross-sections into $\gamma\gamma$
    and $e^+e^-$ for purely axial coupling ($a=0$, plain curve) and purely
    scalar ($b=0$, dashed).}
\label{fig_eta}
\end{figure}


The $\gamma\gamma$ annihilation cross-section is then also determined by
this measurement in terms of the ratio of annihilation branching ratios:
\begin{eqnarray}
\label{def_eta}
&&\eta\doteq\frac{\sigma_{\gamma \gamma}}{\sigma_{511}} = \frac{
\alpha^2}{2\pi^2 \, \beta_e}\ \frac{m_e^2}{m_{dm}^2} \times\\
&&  \hskip 5em
\frac{a^2|1+ 2\,(m_e^2-m_{dm}^2)C_0 |^2 + b^2 |2 m_{dm}^2 C_0|^2 }
 {a^2\beta_e^2+ b^2}
\nonumber
\end{eqnarray}
As announced, this ratio plotted in fig.\ref{fig_eta} cannot vanish, whatever the value of $a/b$, so that a
minimum $\gamma\gamma$ flux is guaranteed. As $m_{dm}$ approaches $m_e$ from above, the ratio increases like
$\beta_e^{-3}$ for a pure scalar coupling ($b=0$) and like $\beta_e^{-1}$ for an axial one  ($a=0$). Playing with
this enhancement is however unnatural. It can spoil nucleosynthesis during the LDM primordial annihilations, and
can furthermore be shadowed by the tail of the 511 line. In the table below, we give typical values of the ratio
$\eta$ for the most conservative case (i.e. $a=0$, $\beta_e^{-1}$). The annihilation cross section into two
photons decreases almost linearly with the dark mater mass for $\mdm > 1$ MeV:
$$
\begin{array}{|r|cccc|}
\hline
m_{dm}(\textrm{MeV}):&  0.52 & 1 & 5 & 20
\nonumber\\
\hline
\eta(a=0):& 8.8 \, 10^{-5} & 1.4 \, 10^{-5}& 3.6 \, 10^{-6} &
8.1 \, 10^{-7}\\
\hline
\end{array}
\label{table1}
$$

Finally, let us notice that the above computation applies equally well to late decaying (instead of
pair-annihilating) dark matter models, such as the intriguing unification of dark matter and dark energy in a
single modulus field proposed in Ref.~\cite{yanagida,Takahashi:2005kp}. The discussion in the next section will
therefore concern both annihilating and decaying dark matter.

In the case of decaying particles, the angular distribution of 511 keV radiation traces the dark matter density
$\rho_{dm}$ (instead of $\rho_{dm}^2$). The latter needs to increase at least twice as fast as the annihilating DM
halo profile towards the galactic center, {\it e.g.} $\sim r^{-2}$ instead of $r^{-1}$, to provide a good fit to
SPI data. Although probably too aggressive, such an increase may not be totally excluded. As pointed out in
Ref.~\cite{Kasuya:2006kj}, it would then be interesting to look for the monochromatic $\gamma$'s produced by the
same decaying dark matter as that which would produce the positrons inducing the 511 keV line. The simple guess
used in Ref.~\cite{Kasuya:2006kj}
$$ \eta_{guess}\approx \frac{\alpha^2 m_{dm}^2}{2\pi^2 m_e^2\beta_e^3}$$
however increases instead of decreasing with $\mdm$. For a typical mass of 10 MeV, this guess overestimates the
monochromatic flux by a factor 635 with respect to our result (Eq.\ref{def_eta}). As we will see in the next
section, such a factor is crucial to the line observability.

\section{Detectability of the monochromatic line}

A few experiments have already scanned the energy range above the electron mass. The instruments on board of
INTEGRAL for example have been designed to survey point-like objects as well as extended sources over an energy
range between 15 keV-10 MeV. The instrument INTEGRAL/SPI itself is a spectrometer designed to monitor the 20 keV-8
MeV range with excellent energy resolution.  Therefore a legitimate question is whether or not the line
$E_\gamma=m_{dm}$ could have been (or could be) detected by the same instrument that has unveiled the 511 keV
signal. This essentially depends on the ratio $\eta$ as given in Fig.~\ref{fig_eta}, and on the background.

%
\begin{figure}
  \includegraphics[width=7cm]{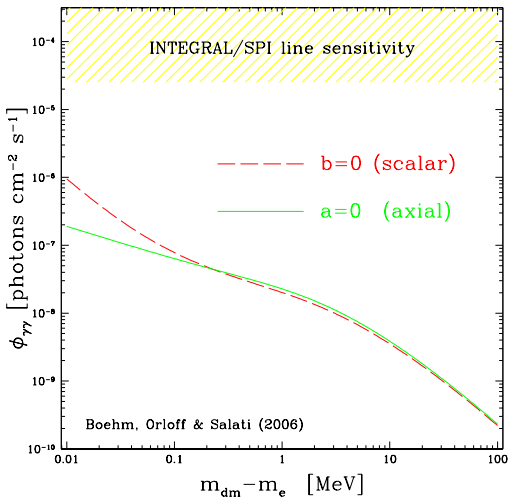}
  \caption{Flux of the monochromatic $E_\gamma=m_{dm}$ line from a 8
    degree cone around the galactic center.}
\label{fluxgg}
\end{figure}

%

The 511 keV emission has been measured with a $\sim$ 10\% precision
\cite{Strong:2005zx} to be
$$
\langle I_{511} \rangle = 6.62 \times 10^{-3} \;
{\rm ph \; cm^{-2} \; s^{-1} \; sr^{-1}}
$$
inside a region that extends over $350^\circ < l < 10^\circ$ in longitude
and $|b| < 10^\circ$ in latitude. If this emission originates from a NFW
distribution of LDM species around the galactic center with a characteristic
halo radius of 16.7 kpc, the signal from the inner $5^\circ$ is found
\cite{ascasibar} to be
$$
\langle I_{511}(5^\circ) \rangle = 1.8 \times 10^{-2} \;
{\rm ph \; cm^{-2} \; s^{-1} \; sr^{-1}} \;\; ,
$$
once the SPI response function is taken into account and the instrumental background is properly modeled. If the
positron propagation is negligible, then the map of the 511 keV emission should correspond to that of the LDM
annihilations.

Within this approximation, we expect the spatial distributions of both the 511 keV and the two-gamma ray lines to
be identical and their intensities to be related by the ratio $\eta$:
$$
\langle I_{\gamma \gamma}(\theta_{\gamma \gamma}) \rangle = \eta \; \frac{\theta_{511}}{\theta_{\gamma \gamma}}\;
\frac{\langle I_{511}(\theta_{511}) \rangle}{(1-3f/4)} \;\; .
$$
This expression is approximately valid as long as the angular radii $\theta_{\gamma \gamma}$ and $\theta_{511}$ of
the regions monitored by the gamma-ray spectrometer are small\footnote{A useful approximation of the intensity for
an NFW profile with $r_0=16.7$ kpc is given by:
$$ I(\theta) = \frac{I_0}{\theta\ [1 + (\theta/11^\circ)^{1.07} ]}.$$
This expression is accurate to 5$\%$ for $\theta < 120^\circ$ and shows that deviations from $\rho \sim r^{-1}
\Leftrightarrow I(\theta)\sim \theta^{-1}$ only appear for $\theta > 10^\circ$.}. In what follows, the fraction
$f$ of positrons forming positronium has been taken equal to $93 \%$ as in Ref.~\cite{ascasibar}. This is in
perfect agreement with the positronium fraction derived in Ref.~\cite{Weidenspointner:2006nu}, i.e. $f_{Ps} = 0.92
\pm 0.09$. The monochromatic line flux
$$
\phi_{\gamma \gamma}(< \theta_{\gamma \gamma}) = \pi \, \theta_{\gamma \gamma}^{2} \;
\langle I_{\gamma \gamma}(\theta_{\gamma \gamma}) \rangle
$$
has been plotted in Fig.~\ref{fluxgg} in the case of the LDM model with $F$ exchange and assuming a NFW profile.
The angular radius $\theta_{\gamma \gamma} = 8^\circ$ corresponds to the field of view of the satellite. For
typical LDM masses in the MeV range, the expected flux is about three orders of magnitude below the claimed
INTEGRAL/SPI line sensitivity \cite{Roques:2003xg} (which is about $2.5 \times 10^{-5}$ ph cm$^{-2}$ s$^{-1}$
after $10^{6}$ seconds). An unrealistic exposure of 30,000 years would thus be required in order to detect the $E=
\mdm$ line. When the LDM species is degenerate in mass with the electron, the flux is only a factor of 25 below
the SPI detection limit (assuming a pure scalar coupling b=0). As long as the mass difference $m_{dm} - m_{e}$
does not exceed 0.1 MeV, it is roughly comparable with the expected 478 keV line signal emitted by Novae
\cite{Teegarden:2006ni}, that is about $\sim 10^{-7}$ ph cm$^{-2}$ s$^{-1}$.

\begin{figure}
  \includegraphics[width=7cm]{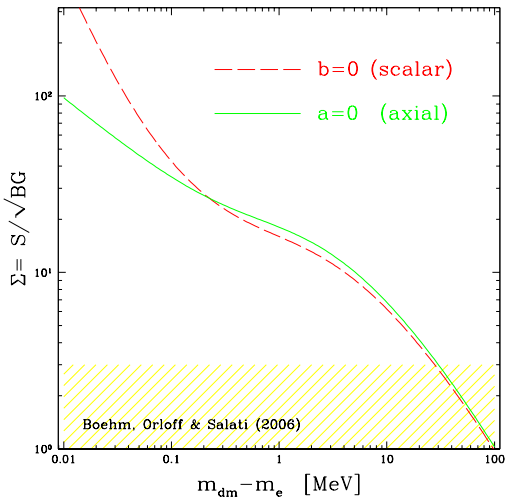}
  \caption{Significance of the monochromatic $E_\gamma=m_{dm}$ line above the
    continuum background for one year of observation with an ideal detector of
    1 m$^{2}$ and a $10^{-3}$ energy resolution.}
\label{sigmagg}
\end{figure}


 SPI sensitivity is limited by the instrumental background that arises
mostly from cosmic rays impinging on the apparatus and activating the BGO scintillator. On the contrary, the
absolute sensitivity of an ideal instrument is purely limited by the gamma-ray continuum background. This emission
has been recently estimated by Ref.~\cite{Strong:2005zx} who found
$$
I_{BG}(E) = 1.15 \times 10^{-2} \, E^{-1.82} \;
{\rm ph \; cm^{-2} \; s^{-1} \; sr^{-1} \; MeV^{-1}} \;\; ,
$$
inside the central region that extends over $350^\circ < l < 10^\circ$ in
longitude and $|b| < 10^\circ$ in latitude. The energy $E$ is expressed
in units of MeV.

We thus estimate the significance
$\Sigma \equiv \textrm{signal} / \sqrt{\textrm{background}}$ for the LDM line
to emerge above this background (assuming it is isotropic) to be
$$
\Sigma = \sqrt{\pi} \; \theta_{511} \;
\frac{\langle I_{511}(\theta_{511}) \rangle}{(1-3f/4)} \; \eta \;
\sqrt{\frac{S_0 \, T_0}{I_{BG} \, \Delta E_0}} \;\; ,
$$
with $S_0$ the surface of the detector, $T_0$ the exposure time, $I_{BG}$ the above-mentionned continuum
background intensity and $\Delta E_0$ the energy resolution. The significance $\Sigma$ (displayed as a function of
the dark particle mass in Fig.~\ref{sigmagg} for a surface of 1 m$^{2}$, an exposure duration of $T_0 = 1$ year
and an energy resolution of $0.1 \%$) indicates that those values would theoretically allow to extract the minimal
guaranteed signal computed at 3 standard deviations above background for all relevant LDM masses below 30 MeV.

There is nothing to be gained by narrowing the angular aperture $\theta_{\gamma \gamma}$ because, for the assumed
NFW profile, the signal increases linearly with this angular radius, as does the square root of an isotropic
background.

In contrast, note that the monochromatic line should be extremely narrow: its width is expected to be about a few
eV which experimentally is very challenging if one compares it with the present SPI sensitivity that is about
$10^{-3}$ at MeV energies.  At lower energies, there are nevertheless instruments, e.g. X-ray CCD, bolometers,
Bragg spectrometers which are able to resolve eV widths. A significant improvement on the resolution $\Delta E_0$
at higher energies would probably be necessary in order to reach a large enough significance and ensure detection.
Indeed, an effective surface of 1m$^{2}$ might be hard to attain in space.

\section{conclusion}

In this paper, we computed the pair annihilation of LDM particles into two photons $\sigma_{\gamma \gamma} v_r$
and determined the flux $\phi_{\gamma \gamma}$ associated with the monochromatic line $E=m_{dm}$.  To obtain a
conservative estimate, we considered only $e-F_e-\rm{dm}$ interactions and ignored all other possible
interactions. With this simplistic assumption we could relate our estimate of $\phi_{\gamma \gamma}$ to the 511
keV flux that has been measured by INTEGRAL/SPI. We made the reasonable assumption that the particle $F_e$ was
much heavier than the dark matter and the electron. We also assumed that the couplings were small enough
($(c_l^2+c_r^2) m_e \ll 2 c_l c_r m_{F_e}$) so that the contribution associated with the electron mass in the
cross section could be neglected.

We found that $\phi_{\gamma \gamma}$ was ranging from $10^{-6}$ to $10^{-10} \ {\rm ph \; cm^{-2} \; s^{-1}}$ for
dark matter masses from $m_e$ to 100 MeV. These values are well below the present SPI sensitivity.

Next generation instruments such as AGILE/(super AGILE) or GLAST,
which in principle could be more promising, will probably be limited
by the energy range that they are able to investigate. Future
instruments might nevertheless be able to see this line if their
energy resolution and sensitivity are improved by a large factor
with respect to SPI present characteristics.

Maybe a better chance to detect this line is to do observations at a high latitude and a longitude slightly off
the galactic centre. In this case, indeed, the background should drop significantly (the density of dark clouds
has been measured recently \cite{grenier}) but the line flux may decrease by a smaller factor.

In dwarf galaxies,  where the dark matter content dominates over baryons, the gamma ray background is also
expected to be quite suppressed. The line $E=m_{dm}$ might be easier to detect.

Among the closest dwarfs to us, Sagittarius Dwarf Galaxy (located at a distance of 24 kpc from us and with a size
of about $10^8 \ M_{\odot}$ \cite{Ibata:1995fz}), is particularly interesting. The amount of intrastellar gas is
very low and the dwarf contains a large amount of popII stars. The problem is that it is somehow hidden by the
galactic centre although it is a bit off.

In principle, dwarf spheroidals are a powerful tool for testing the
LDM hypothesis. E.g. the detection of a bright 511 keV line within
INTEGRAL's sensitivity would provide a strong confirmation of this
scenario \cite{hooper-dwarf}. However, its detectability relies on
the hypothesis that there are enough electrons to thermalize and to
stop the positrons. No gas has ever been detected in any of the
local group dwarf galaxies so it is hard to make reliable estimates.
Also one needs to know the spatial distribution of the gas to make
accurate predictions and determine whether the 511 keV emission will
be extended or not.

SDG is being disrupted by the tidal forces of our galaxy so the approximation of a spherical DM halo profile
probably leads to incorrect predictions.

At last, depending on the fraction of gas, inflight $e^+ e^-$ annihilations may happen before the positrons have
time to thermalize. In this case, there should be a broad line at an energy $m_e < E < m_{dm}$\footnote{Note that,
for the Milky Way, the detection of both the 511 keV line and the continuum indicates that the interstellar medium
is in a warm partially ionized phase. However the temperature of the gas in the dwarf may be larger. Even if the
positrons can annihilate after being thermalized, this would imply a suppression of the positronium formation rate
and would favour inflight $e^+ e^-$ annihilations.} instead of a 511 keV line.

The study of the $\gamma \gamma$ channel has therefore two advantages compared to the 511 keV emission: it does
not rely on the estimate of the electron number density nor the gas fraction inside the dwarf. It is also
independent of the gas spatial distribution and can probe directly the DM halo profile. It has one major drawback:
it is a higher order process and it is therefore suppressed compared to the the $e^+ e^-$ production. Since
Ref.~\cite{cordier} looked for the 511 keV line in SDG and did not find it, the chance to detect the monochromatic
line is probably very small although Ref.~\cite{cordier} was unable to probe the upper limit of the predicted
range and the number density of electrons inside the dwarfs may have been overestimated.

\section*{Acknowledgement}
We are grateful to Y. Ascasibar, M. Casse, ISDC members, E. Pilon and A.W. Strong, for helpful comments. The
authors thank the PNC (french Programme National de Cosmologie) for financial support to this work.

\bibliography{line}

\end{document}